\documentclass[lettersize,journal]{IEEEtran}
\usepackage{amsmath,amsfonts}
\usepackage{algorithmic}
\usepackage{algorithm}
\usepackage{array}
\usepackage[caption=false,font=normalsize,labelfont=sf,textfont=sf]{subfig}
\usepackage{textcomp}
\usepackage{color,soul} 
\usepackage{stfloats}
\usepackage{url}
\usepackage{verbatim}
\usepackage{graphicx}
\usepackage{cite}
\usepackage{booktabs}
\usepackage{tabularx}
\usepackage{caption}
\usepackage{lipsum}
\usepackage{pdflscape}
\usepackage{tabularx}
\usepackage{longtable}
\usepackage{tabularx}
\usepackage{booktabs}
\usepackage{caption}
\usepackage{stfloats}
\usepackage{hyperref}
\usepackage[nolist,nohyperlinks]{acronym}
\begin{acronym}
\acro{HRI}{human-robot interaction}
\acro{AI}{artificial intelligence}
\acro{GUI}{Graphical User Interface}
\acro{SDK}{Software Development Kit}
\acro{ROS}{Robot Operating System}
\acro{CAD}{Computer-Aided Design}
\acro{API}{Application Programming Interface}
\acro{RQ}{Research Questions}
\end{acronym}

\IEEEoverridecommandlockouts 

\AtBeginDocument{%
  }
    \small 

\begin{document}

%\title{FONTANA – Fostering rObotics iNvolvemenT for Artistic maNufActs}
\title{Artists' Views on Robotics Involvement in Painting Productions}

\author{
Francesca Cocchella$^{1}$, 
Nilay Roy Choudhury$^{2}$, 
Eric Chen$^{2}$, 
and Patrícia Alves-Oliveira$^{2}$\\[0.5em]
\small $^{1}$CONTACT Unit, Italian Institute of Technology, Genoa, Italy\\
\small $^{2}$University of Michigan, Ann Arbor, USA
}
        % <-this % stops a space
%\thanks{This paper was produced by the IEEE Publication Technology Group. They are in Piscataway, NJ.}% <-this % stops a space
%\thanks{Manuscript received April 19, 2021; revised August 16, 2021.}}



% \begin{verbatim}
%   \begin{teaserfigure}
%     \includegraphics[width=\textwidth]{Figures/artworks.png}
%     \caption{figure caption}
%     \Description{figure description}
%   \end{teaserfigure}
% \end{verbatim}
\maketitle

\begin{abstract}
As robotic technologies evolve, their potential in artistic creation becomes an increasingly relevant topic of inquiry. This study explores how professional abstract artists perceive and experience co-creative interactions with an autonomous painting robotic arm. Eight artists engaged in six painting sessions—three with a human partner, followed by three with the robot—and subsequently participated in semi-structured interviews analyzed through reflexive thematic analysis. Human-human interactions were described as intuitive, dialogic, and emotionally engaging, whereas human-robot sessions felt more playful and reflective, offering greater autonomy and prompting for novel strategies to overcome the system's limitations. This work offers one of the first empirical investigations into artists' lived experiences with a robot, highlighting the value of long-term engagement and a multidisciplinary approach to human-robot co-creation.
\end{abstract}


%   \includegraphics[width=\textwidth]{Figures/artworks.png}

\section{Introduction}

\IEEEPARstart{A}rt has always been a mirror of the human spirit. It is an intimate expression of emotion, identity, and cultural context. Meanwhile, technology is becoming increasingly intertwined with our daily lives, and it is beginning to shape how we create, perceive, and share artistic experiences. If art reflects who we are, then its process evolves as technology shapes how we live and create.
Recently, robotic systems have been employed as creative agents. From algorithmically-generated paintings \cite{schaldenbrand2022frida} to autonomous robotic brushstrokes  \cite{schurmann2025neuromorphic}, these technologies challenge traditional boundaries between the artist and the medium. While research has focused on what robots can contribute to art, little is known about how artists experience creating \textit{with} them.
%Likewise, how such hybrid artworks are interpreted and valued by art-world professionals remains an open question.

\begin{figure}[t] 
    \centering  
    \includegraphics[width=\linewidth]{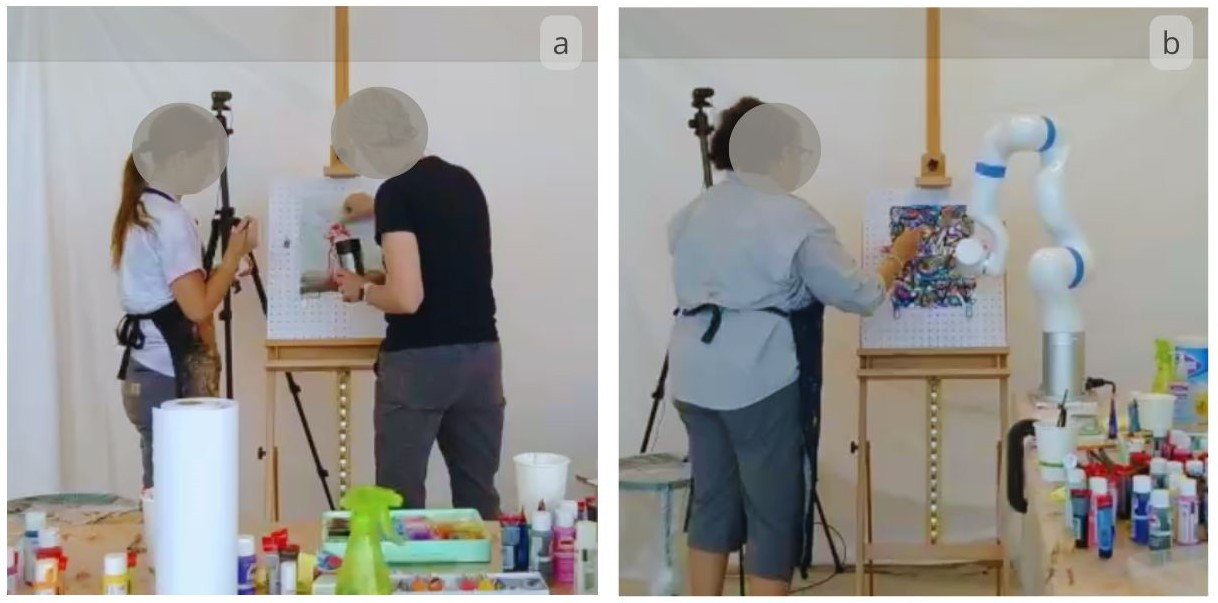} 
    \caption{\small{Study conditions: (a) artist-artist collaborative painting; (b) artist-robot collaborative painting.}  }
    \label{fig:conditions}  
\end{figure}

We examine how artists experience collaboration with a robot during artistic creations.
%, positioning them as key stakeholders in human–robot co-creation.
%As machines begin to take part in artistic creation, once considered a uniquely human domain, we are prompted to rethink what it means to create and to collaborate. 
%This rethinking requires attention not only to the output of H-R collaboration but to the lived and situated dynamics that shape the artistic process.
To achieve this,
%explore the role of robots in artistic creation from expert perspectives,
we conducted a longitudinal study where artists and the robot performed collaborative paintings. Each artist performed a total of $6$ painting sessions, $3$ with another human artist and $3$ with a robot (Figure \ref{fig:conditions}). Through semi-structured exit interviews, we explored their reflections on the nature of the interaction, the perceived agency attributed to the robot, the creative flow, and how they perceived the robot's ability to engage socially and contribute artistically to the paintings.
Our findings advance understanding of human–robot collaboration in the arts by foregrounding artists’ lived experience.
%and the cultural reception of robot-assisted creation.}
%--- with another human artist and with a robot. Using diary entries, video recordings of the interactions, and semi-structured interviews, we have explored how artists interpret and engage with the creative agency of the robot. 

% \begin{figure}[t] 
%     \centering 
%     \includegraphics[width=\linewidth]{Fig/zine-template.png}  
%     \caption{Zine template used in the zine-making workshop described in Section~\ref{sec:zine-making}.}  
%     \label{fig:zine-template}  
% \end{figure}

\section{Related Works}
%In this section, we briefly discuss the concept of collaboration in human interactions. We then shift focus to explore collaboration between robots and humans in artistic contexts. Finally, we address the importance of stakeholders' involvement in the development and evaluation of a collaboration with a robotic platform.

\subsection{Human-Human Artistic Collaboration}
Collaborative painting between humans offers a rich context to explore social dynamics, such as co-presence, mutual adaptation, and interpersonal synchrony. Prior work has shown that such interactions are often experienced as fluid, dialogic, and emotionally engaging, supported by verbal and non-verbal coordination~\cite{ rabinowitch2017synchronized}.
Abraham et al.~\cite{abraham2023painting} extended this understanding by demonstrating that even in the absence of rhythmic structure, visual coordination during joint painting can foster perceptions of harmony, empathy, and mutual understanding.
%Their findings highlight how co-creation in the visual arts can function as a powerful social signal, shaping how collaborators perceive one another and their shared work.
This body of work provides a valuable foundation for interpreting how artists in our study experienced human collaboration. Prior work on human–human artistic collaboration highlights features such as mutual adaptation, interpersonal synchrony, and continuous negotiation. An open question is whether—and in what form—these features may also emerge when artists collaborate with a robot.

\subsection{Human-Robot Artistic Collaboration}
The intersection of robotics and art has been a complex domain for exploring creativity and the evolving role of machines as collaborators or creators. Both early and contemporary works illustrate a wide range of approaches to human–robot artistic collaboration, including painting and other forms of artistic practices.
%beyond painting, with painting representing a particularly rich and long-standing area of exploration.}
For instance, \textit{Telegarden} explored the possibilities of telerobotic drawing and cloud-mediated interaction~\cite{goldberg2001telegarden}, while the \textit{e-David} project—developed in collaboration with artist Liat Grayver—demonstrated the use of semi-autonomous systems in iterative painting processes~\cite{gulzow2018self}. Other artists, such as Leonel Moura, have investigated autonomous swarm-based robotic painting~\cite{moura2018robot}, whereas Sougwen Chung emphasized the co-creative potential of \ac{HRI} in real-time artistic production~\cite{jane2023machines}.
%\\Collectively, these efforts illustrate the various ways in which robotic systems can participate in artistic production, ranging from tools and performers to genuine co‑creators.

Recent research in robotic art has moved toward approaches that highlight the creative process itself, treating the making of the artwork as equally important as the finished piece.
%Recent research has pushed the field toward process‑oriented and expressive forms of robotic art, where the emphasis is as much on the act of creation as on the final product.
Schürmann et al. ~\cite{schurmann2025neuromorphic} demonstrated that robotic brushstrokes can capture the variability and spontaneity of human movement, emphasizing
%adaptive and materially grounded
co‑creation rather than strict replication.
%Their work shows how incorporating physical media and real‑time feedback can introduce natural variation that enhances expressiveness. 
Additional work on this domain includes the study of creative practices by Qin et al.~\cite{qin2025encountering}, and the work of the artist Melkio who used joysticks to teleoperate the humanoid robot in the creation of abstract paintings~\cite{melkioRobot}.
% Qin et al.~\cite{qin2025encountering} studied the creative practices of professional artists using robots, highlighting that creativity emerged from the interplay of human input, environmental context, and the acceptance of uncertainty as a generative force. Another example is the Human + Robot Painting Project ~\cite{melkioRobot},
% %\footnote{Human + Robot Painting Project:\url{https://melkio.com/robotmelkio/}}
% where pop surrealist artist Melkio used joysticks to teleoperate the humanoid robot Alter-Ego in the creation of abstract paintings.
% %The artist likened the experience to painting through an instinctive, childlike intermediary.
Taken together, these works reflect a shift from pure automation toward iterative, human‑in‑the‑loop creations.

Peter Schaldenbrand, Gaurav Parmar, Jun-Yan Zhu, James McCann, and Jean Oh's work on the FRIDA and COFRIDA robot systems introduced methods for robotic painting that leverage style transfer and generative modeling to produce novel artworks in collaboration with humans~\cite{schaldenbrand2022frida, schaldenbrand2024cofrida}. Their approach frames the robot as a co‑creative partner capable of stylistic adaptation and expressive output through discrete turn-based painting. Our robot painting system builds on this work by extending it toward more collaborative, iterative painting, where the robot acts as an independent creative partner. This allowed us to explore its perceived role, agency, and presence within the shared artistic process.

\subsection{Stakeholders' Perceptions of Robots}
Understanding how people perceive and engage with robots can guide the development of systems that support creative expression and collaboration~\cite{gruneberg2020culturally}.
%Well-designed user studies offer valuable insights into how stakeholders experience emerging technologies.
Increasingly, research emphasizes the importance of involving stakeholders early
%and adopting bottom-up approaches
in robotics development~\cite{soraa2023older}. Stakeholders differ from end-users in that they are directly impacted by the technology's role and purpose, not just its functionality, promoting acceptance %and meaningful integration
by addressing the social dynamics inherent to \ac{HRI}~\cite{fraune2022lessons}. 

Recent studies have begun to explore how artists, i.e., stakeholders, perceive and engage with emerging technologies. A qualitative study with expert avatar designers revealed tensions around treating \ac{AI} as a co-creator and managing its iterative outputs~\cite{he2023exploring}. %While focused on disembodied systems, these insights can inform embodied robotics by emphasizing how physical presence and interaction may further support creative empowerment and expression.
Qin et al.~\cite{qin2025encountering} investigated the creative practices of professional robotic artists, showing how creativity emerges from the interplay of human input, context, and uncertainty. Their use of semi-structured interviews enabled a deep exploration of the experiences, illustrating the value of these methods in stakeholder research.

Qualitative methods, and semi-structured interviews in particular, are especially effective as they offer the flexibility to adapt to stakeholders' narratives while enabling researchers to access detailed, context-specific insights.
%from individuals with domain-specific expertise.
This approach aligns with our goal of understanding how stakeholders make sense of creative \ac{HRI}, which is the focus of this research.
%Importantly, it is necessary to distinguish between end-users and stakeholders. End-users are those who interact with a system directly in its final form, while stakeholders—such as professional artists in our study—play a broader role in shaping, evaluating, and attributing meaning to the technology. 

\section{Our Contribution}
This study investigates how professional artists, i.e., stakeholders, perceive the involvement of robotics in painting productions, by comparing their experiences of collaborating with a human and with an autonomous robot. Through painting sessions and exit interviews, we gathered stakeholders' insights to inform the future design of human-robot artistic collaboration.

\section{System Overview}
\begin{figure}[t]
    \centering\includegraphics[width=\linewidth]{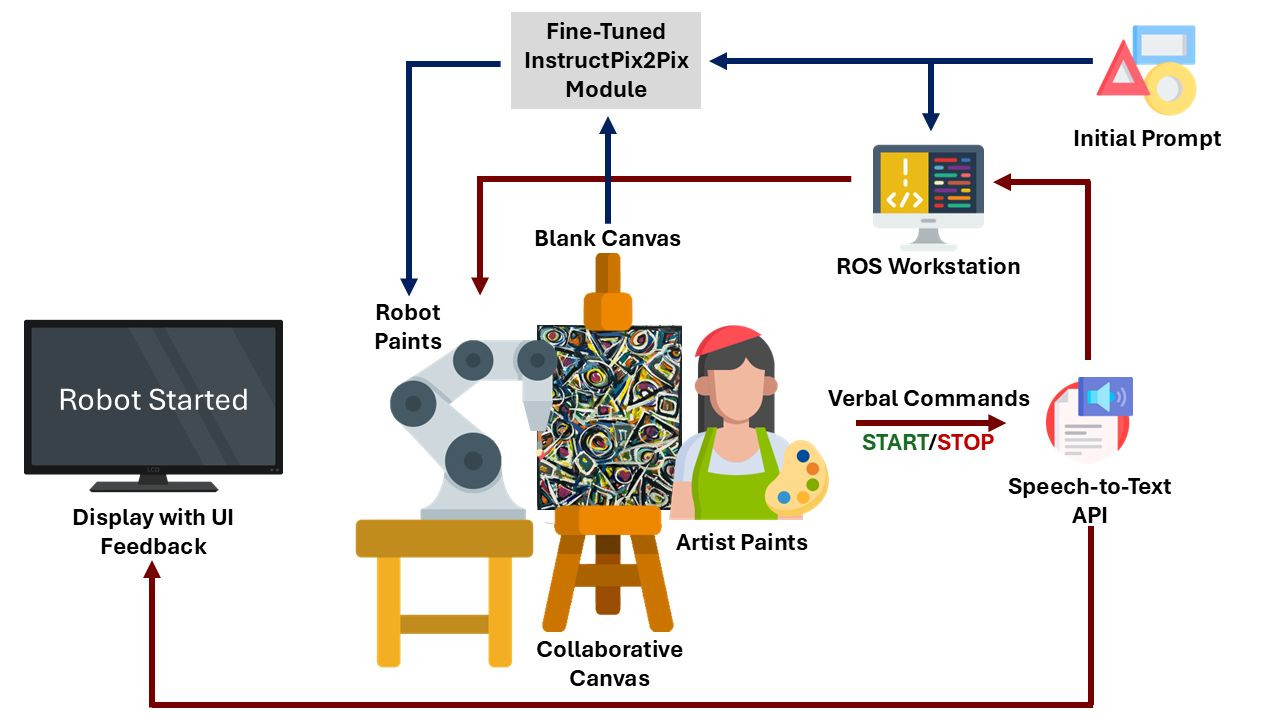}
    \caption{\small{System architecture of the robotic painting platform.}}
    \label{fig:system}
\end{figure}

\subsection{Hardware Setup}
\begin{figure}[t] 
    \centering  % Center the image
    \includegraphics[width=\linewidth]{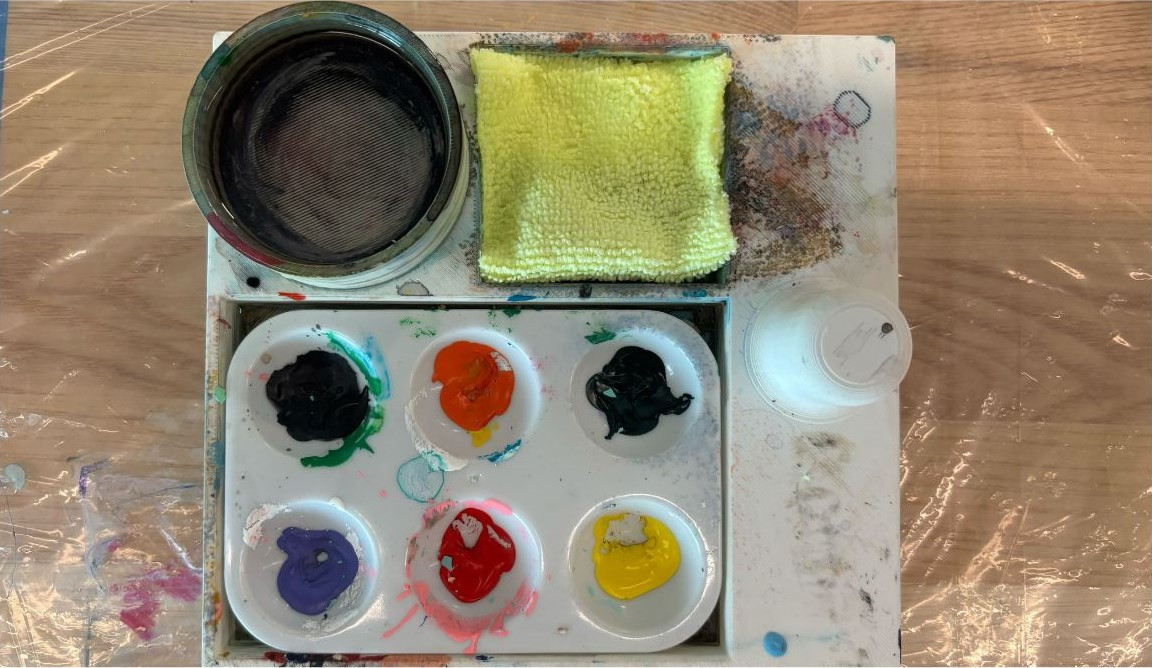}  
    \caption{\small{Close-up on robot painting tools displayed at the table: color palette, water container, and cleaning rag are each mounted on fixed, 3D-printed holders designed for easy and autonomous robot access.} } 
    \label{fig:robot-tools}  
\end{figure}

We now describe the robot system of the robotic platform implemented for this study (see Figure~\ref{fig:system}).
The system utilizes a UF850 6-DOF robotic arm  %\footnote{\url{https://www.ufactory.us/product/850}}
positioned in front of a canvas. The palette, water container, and cleaning rag are each mounted on fixed, 3D-printed holders designed specifically for easy access and repeatable positioning (see Figure~\ref{fig:robot-tools}). These components are arranged to optimize reachability while minimizing the chance of collision during painting operations. %Our hardware setup also includes a custom-designed, spring-loaded brush attachment described in Section~\ref{brush}. 
%Inspired by the cam and latch mechanism of ballpoint pens \cite{us3205863A}, this design allows the robot to swap brushes using a single press-and-release motion. The brush tip is mounted on a retractable shaft with an internal latching system that holds it in place when extended and releases it under axial force. This enables the artist to effortlessly engage and disengage brushes during a session.

\subsection{ROS 2 Node Architecture}
The robotic painting system was developed using the \ac{ROS} 2 framework to interface with the UF850 robotic arm. The core \ac{ROS} 2 node architecture was deliberately kept minimal to improve modularity and clarity for future extensions. The node architecture integrates the \texttt{rclpy} client library and leverages the XArm Python \ac{SDK} for low-level control of the robot. The strokes are mapped to specific robot actions through state transitions. This allows an artist to engage with the system intuitively, without needing to interact with complex terminal commands or \ac{GUI}s. The architecture also allows access to useful robot diagnostics such as joint angles, end effector pose, motion state, and torque using service calls and utility functions from the XArm \ac{SDK}. This supports real-time monitoring and debugging via tools like PlotJuggler and RViz.

\subsection{State Machine Architecture}
A purposely streamlined \ac{ROS} 2-based state machine was developed to govern the robot's behavior. The architecture was designed with only the essential states needed for this study, ensuring clear operation and seamless integration with the artistic interaction while preserving the flexibility for more complex workflows in future iterations. The state machine supports the following key states:

\begin{itemize}
    \item \texttt{IDLE} – Default state where the robot waits for input.
    \item \texttt{GO\_HOME} – Commands the robot to return to a default home pose.
    \item \texttt{PAINTING} – Moves the robot into canvas mode to begin or resume painting.
    \item \texttt{CHANGE\_PAINT} – Moves the robot to the palette region to switch colors or clean the paint brush.
\end{itemize}

In this study, only two voice commands were actively used: \texttt{start} and \texttt{pause}, allowing the artist to initiate or halt the robot's painting actions with minimal cognitive load. %without interrupting the creative flow.

\subsection{Vertical Easel Support}
In this study, the robot painted in a vertical easel showed in Figure \ref{vertical}(a). To enable the robot to paint effectively, a coordinate transformation was implemented to map on-table canvas coordinates to on-easel canvas coordinates. Before runtime, the center position and orientation of the easel in the robot’s base frame are manually calibrated. During operation, points specified in the flat, screen-based coordinate system are first scaled to match the physical dimensions of the easel canvas. These points are then rotated and translated to align with the calibrated easel frame. This transformation ensures that drawings defined in a top-down frame are accurately projected onto the vertical surface, allowing the robot to execute painting motions aligned with user intent.

\subsection{Brush Swapping Mechanism} \label{brush}
\begin{figure}[t] 
    \centering  % Center the image
    \includegraphics[width=\linewidth]{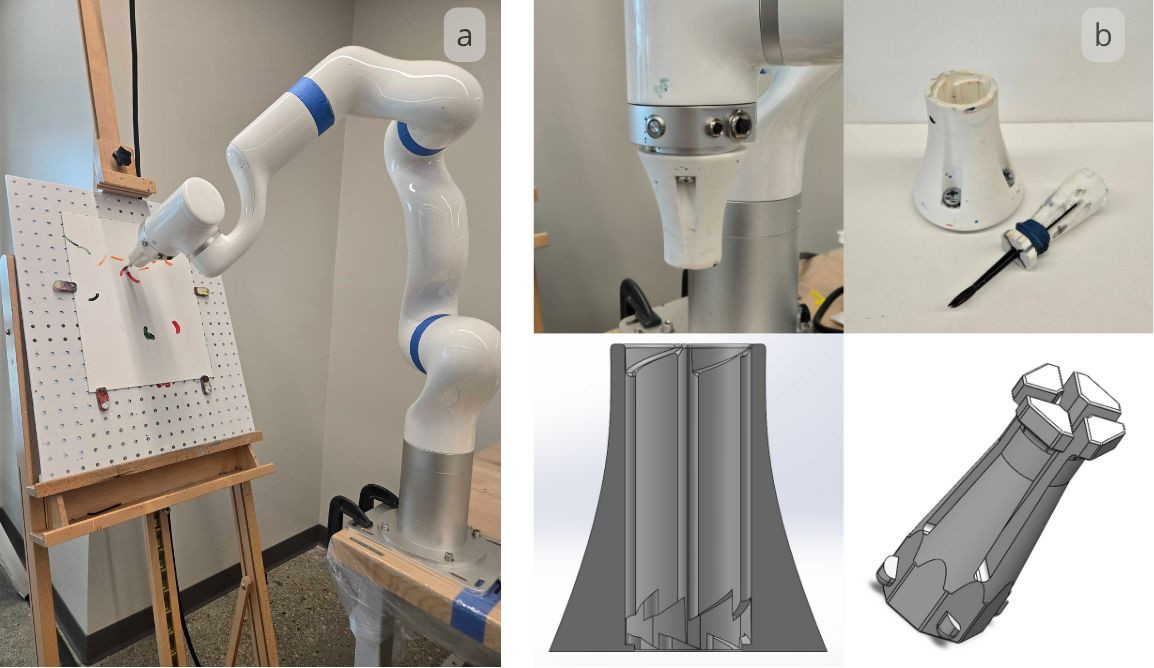}  
    \caption{\small{(a) Robot painting on vertical easel with a fixed canvas size of 11x14in / 27.94x35.56cm; (b) Brush swapping mechanism. Top-left: mounted brush holder on the UF850 robotic arm. Top-right: main body and brush collet with a paintbrush. Bottom-left: internal cam-slotted geometry of the holder. Bottom-right: 3D model of the flexible brush collet designed for tool-free insertion and removal.}}  
    \label{vertical}  
\end{figure}

We developed a custom-designed brush-swapping mechanism through several design iterations. The final version V3.1 is mechanically inspired by the retractable ballpoint pen~\cite{us3205863A} mechanism and allows for quick, tool-free brush changes. The setup consists of two primary components: a \textit{main body} mounted on the robot's end effector and a \textit{brush collet} that securely holds the brush. Figure~\ref{vertical}(b) shows a collage of the final printed mechanism and \ac{CAD} model, highlighting the simplicity and versatility of the design.

The collet is split into four flexible sections and features slanted nubs that allow it to snap into position inside the main body, providing a stable and repeatable interface that can accommodate a range of brush sizes and shapes. Although the robot was not currently programmed to perform autonomous brush changes, the mechanical system is designed with that capability. Currently, the system enabled the artist to easily swap brushes between painting segments by pressing the collet into or out of the main body with a simple push-pull motion.

% \begin{figure}[h]
%     \centering
%     \includegraphics[width=0.3\textwidth]{Figures/brush.png}
%     \caption{Brush swapping mechanism. Top-left: mounted brush holder on the UF850 robotic arm. Top-right: main body and brush collet with a paintbrush. Bottom-left: internal cam-slotted geometry of the holder. Bottom-right: 3D model of the flexible brush collet designed for tool-free insertion and removal.}
%     \label{fig:brush}
% \end{figure}

\subsection{Voice Integration}
The audio subsystem enables real-time voice command recognition for robot control, allowing hands-free natural language communication between the artist and the robot arm. The module is responsible for capturing the audio input, transcribing it to text, and interpreting it into the intended commands for the robot. We used AssemblyAI's \footnote{AssemblyAI: \url{https://www.assemblyai.com/}} transcription \ac{API} for real-time speech-to-text transcription. The audio is captured through a RODE Wireless GO II microphone, only processed at final transcription events at the end of each spoken utterance. Once the complete command is transcribed, a zero-shot classification model labels the user's intent from a predefined set of commands (e.g., \texttt{start}, \texttt{stop}, \texttt{draw}). The commands are then published through \ac{ROS} for the robot to act upon. 

\subsection{Visual Feedback Interface}
To support clarity and accessibility during the interaction, the system included a full-screen interface that displays real-time status updates to the artist. The interface provided visual feedback in response to voice commands, such as \texttt{Robot Starting} and \texttt{Robot Paused} to help the participant confirm if the system has recognized their voice command properly. The interface was built using Python's Tkinter library and runs in parallel with the transcription system through multithreaded execution to ensure interface responsiveness simultaneously with continuous audio streaming and processing. 

\begin{figure}[t] 
    \centering  % Center the image
    \includegraphics[width=\linewidth]{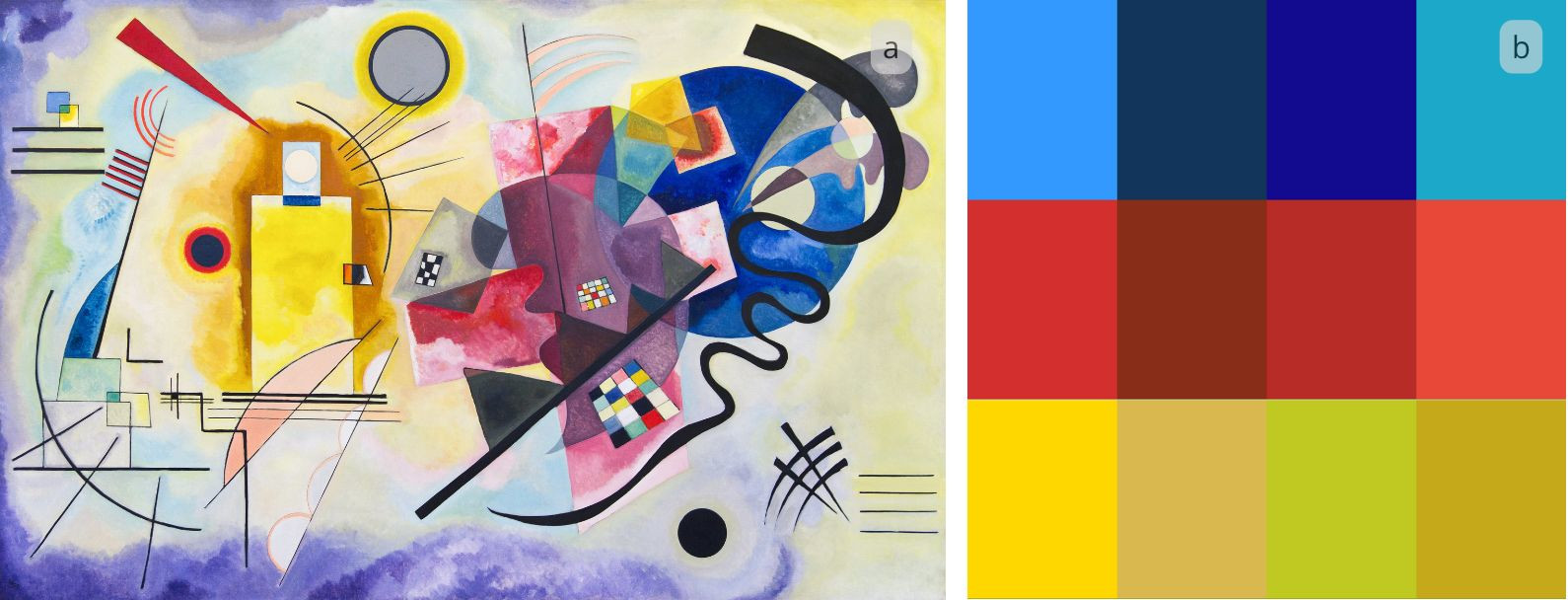}  
    \caption{\small{(a) \textit{Gelb Rot Blau}, Vasilij Vasil'evi\v{c} Kandinskij, 1925. Collection of Centre Pompidou, Paris, France; emblematic example of his color theory linking primary colors with geometric forms, which inspired the experiment set up. (b) Color set-up inspired by Kandinskij's theory for this experiment.}}  
    \label{kand}  
\end{figure}

\subsection{Abstract Dataset and Model}
To support abstract-art painting and styles, we built a custom dataset and fine-tuned an image-translation model tailored to robot painting. We collected a diverse collection of abstract artworks from the Kaggle Abstract Art Gallery \footnote{Kaggle Abstract Art Gallery: \url{https://www.kaggle.com/datasets/bryanb/abstract-art-gallery}}, sourced from WikiArt. We then prompt-engineered GPT‑4 (ChatGPT 4.0) to generate descriptive text labels for each image, creating 300 image-text labels for training. The CoFRIDA framework provided stroke-level painting data by simulating the robot's painting process, generating both partial-progress and completed canvases ~\cite{schaldenbrand2022frida,schaldenbrand2024cofrida}. While CoFRIDA is originally designed to produce paintings that accurately depict their subject matter with coherent stroke patterns and layering, our work adapts this approach specifically for abstract art by fine-tuning the InstructPix2Pix Module originally used within the CoFRIDA framework on a curated dataset of abstract works from Kaggle paired with GPT-4-generated captions. As a result, InstructPix2Pix generates a target image from input text prompts that reflects the desired abstract style and geometric emphasis of a painting session. This target image serves as a high-level visual goal and is subsequently translated into executable robot stroke sequences using the FRIDA image-to-action planning process, which computes stroke trajectories that approximate the target image on the canvas.

 \subsection{Artistic Framework}
 \label{artfram}
\begin{figure}[t] 
    \centering  % Center the image
    \includegraphics[width=\linewidth]{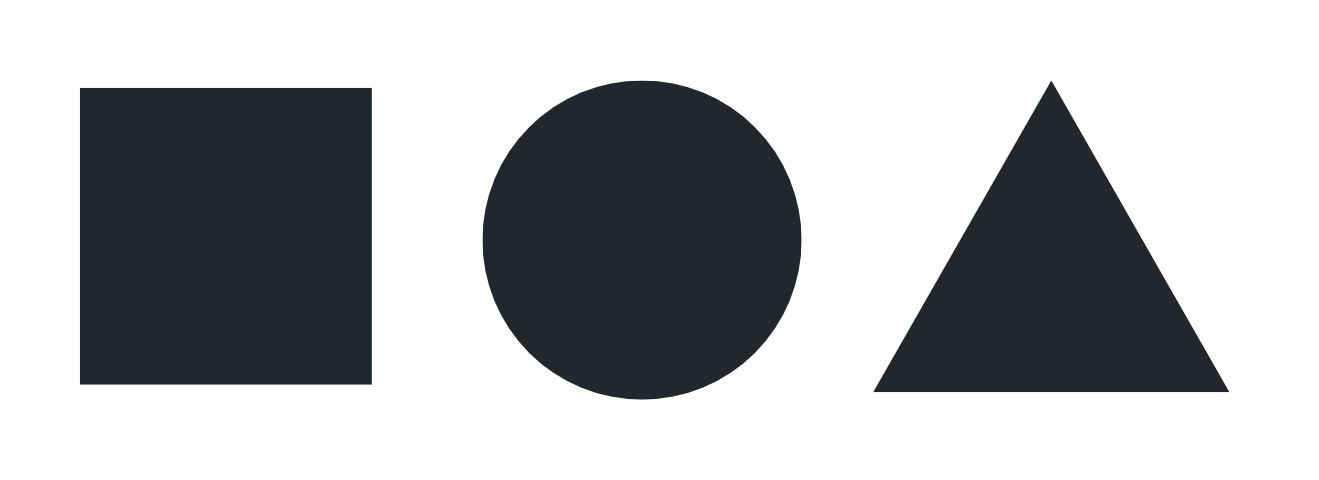}  
    \caption{\small{Studies of geometric shape by Bruno Munari~\cite{munari2015bruno}.} } 
    \label{fig:munari}  
\end{figure}

The robot was programmed to draw circle-inspired shapes in the first session, square-inspired in the second, and triangle-inspired in the third. This setting was inspired by Bruno Munari 's~\cite {munari2015bruno} exploration of geometric forms (see Figure~\ref{fig:munari}). In the 1960s, Italian designer Bruno Munari published visual case studies on Circle, Square, and later Triangle, associating each with specific qualities: the circle with the Divine, the square with safety, and the triangle as a key connective form. \\Associated to this, at the Bauhaus, the Russian-French artist Vasilij Vasil'evi\v{c} Kandinskij explored geometry’s psychological and spiritual effects, viewing the triangle as active and aggressive, and the square as peaceful and calm~\cite{lee1990kandinsky}. 

In our study, we sequenced the robot’s color–shape associations across three painting sessions. Initially, it was given four acrylic colors: blues for the circle, reds for the square, and yellows for the triangle. These associations draw on Kandinskij’s synesthetic theory (Figure \ref{kand}(a) ), linking primary shapes to specific colors, while the focus on forms derives from Munari’s visual design work. The digital versions of the selected colors are shown in Figure~\ref{kand}(b).

% \begin{figure}[H] 
%     \centering  % Center the image
%     \includegraphics[width=0.2
%     \textwidth, scale=0.10]{Figures/color setup.png}  
%     \caption{Color Set Up inspired by Kandinsky's theory for this experiment.}  
%     \label{colset}  
% \end{figure}

\subsection{Stroke Generation and Management}
To support reproducibility and uninterrupted painting sessions, the robot executed pre-generated stroke sequences rather than generating strokes in real time. 

Stroke sequences were generated offline by applying the FRIDA image-to-action planning process as implemented within the CoFRIDA framework to the target images produced by InstructPix2Pix. Given a target image, FRIDA computes a sequence of discrete brush strokes that iteratively reduce the visual difference between the current canvas state and the target image. Each stroke is represented as a short sequence of 3D waypoints in the robot base frame, corresponding to brush approach, paint deposition, and lift-off.

Before human-robot artistic sessions began, we generated strokes based on the artistic framework described in Section ~\ref{artfram} and saved them as serialized session files. For each painting session (circle, square, or triangle), the stroke library contained 330 distinct pre-generated strokes, corresponding to approximately one and a half hours of continuous painting. For each session, we preloaded the corresponding circle, square, or triangle stroke sequences, allowing the robot to immediately begin painting without additional computation. 

Within a session, strokes adhere to the same high-level artistic framework and target image style and are executed deterministically from a pre-generated library, allowing the same stroke sequences to be replayed across multiple participants, thereby maintaining consistent behavior during the human‑robot trials, essential to compare the experiences that artists had with the robot. 

% The stroke library contained enough pre‑generated motions to support painting sessions lasting over an hour without requiring live generative computation or system restarts.

\section{Study Method}

\subsection{Goal and Research Questions}
Our study adopts an exploratory approach and constitutes, to our knowledge, the first attempt to (1) directly compare human–human and human–robot collaborative experiences within an artistic context, and (2) collect artists' reflections and perspectives after engaging in direct interaction with a robot during several painting sessions.

Given the novelty of the topic and the limited existing empirical evidence in this domain, the study does not advance formal hypotheses, but instead is guided by the following \ac{RQ} aimed at uncovering emergent patterns:

\begin{itemize}
    \item RQ1: \textit{ How do artists perceive the experience of collaboratively painting with a robot compared to collaboratively painting with a human?} %We hypothesize that the collaborative dynamics typically observed in human-human artistic collaborations will also emerge in human-robot interactions.
    \item RQ2: \textit{How do artists experience the creative flow with a robot and with a human artist?} %Over the course of the interaction, participants will become more accustomed to the robot, leading from initially rigid dynamics to more fluid and natural collaboration.
    \item RQ3: \textit{What are the artists' perceptions of robot agency during the painting production?} 
    \item RQ4: \textit{What are the artists' perceptions of the robot's capabilities for interaction and painting?}
\end{itemize}

\subsection{Participants}
Participants were recruited through targeted outreach to local artists, painters, and galleries in Ann Arbor and Detroit metropolitan areas. Individuals were asked to complete an eligibility questionnaire through the platform Qualtrics %\footnote{\url{https://www.qualtrics.com/}} 
(available on OSF in Appendix\_a \footnote{Open Science Framework project related to this submission: \url{https://osf.io/yhca7/overview}}), which allowed us to select only professional artists working within the context of abstract art. All participants read and signed an Informed Consent. The research has been approved by the University of Michigan Institutional Review Board (IRB) [HUM$00273497$]. Upon completion of all experimental sessions, participants received a compensation of \$300.
Our study sample included  \textbf{$8$} participants who met our eligibility criteria and agreed to enroll in the study. Participants were paired, balancing for gender and age (see Table~\ref{tab:participants}). From our sample, one participant (A$8$) had previously interacted with a robot, while four artists (A$1$, A$2$, A$3$, and A$8$) reported having collaborated with others in painting before.

%As part of the process, they were also asked to identify one artwork per condition (human and robot) that they considered the final outcome of the collaboration—potentially to be included in a future exhibition. 

\subsection{Procedure}
Our exploratory study employed a within-subjects design where each participant engaged in six painting interactions: three with a human artist (H-H) and three with an autonomous robot (H-R). The order of the sessions was always human-human first and then human-robot. %This design treats learning not as a confound, but as a key aspect of the human-robot co-creative process. 
The sessions were conducted on separate days, with a minimum interval of one day between them to provide artists with time to elaborate and reflect on their experiences before moving to a new session. Overall, the study spanned three weeks. The interactions were held in the University of Michigan. Each interaction required approximately one hour or one and a half hours.

%Measurements and analysis of the data are described below.
%Participants will have the opportunity to keep all artworks produced during the experimental sessions, including sketches, drafts, and exploratory materials. For the final artwork created in collaboration with the human partner, the two participants will decide together who will retain the final piece and associated materials, as suggested in literature [see IRB]. 
%Regardless of ownership, the research team will take high-quality photographs of all materials created during the sessions. These images will be archived and used as part of the study’s experimental data.

Participants were instructed to create one final art piece to represent the outcome of the H-H collaboration, and one final piece to represent the outcome of the H-R collaboration. This final art piece was selected after the last session of each condition. In both conditions, participants were required to collaborate in the artistic product with their partner.
Here, “collaboration” refers to a shared painting process on the same canvas, where human artists and the robot alternately contribute mark-making.
No specific guidelines were provided regarding the type, style, or medium of the artwork, as long as it could be considered abstract art. Participants were free to produce as many pieces as they wished, without limitations. The experimenter was available to clarify any questions about the process but was not present in the room during the creative sessions, to preserve the spontaneity of the interactions.

In the first session, participants were informed they would be provided with painting materials and that they were free to bring any additional supplies they considered necessary for the artwork creation. Furnished painting materials included: a canvas (size: $11 \times 14$ inches / $27.94 \times 35.56$ cm), acrylic and watercolor paints in multiple shades, pencils and markers, brushes of various sizes, glue, and cleaning supplies. Materials brought by the artists included paints of different shades and types (e.g., oil), palette knives, brushes, and stickers.

At the end of each painting session, the research team took high-quality photographs and scans of all materials created by the artists (see Figure~\ref{fig:paintings}). Artists retained ownership of the original pieces. At the end of all six painting sessions, participants took part in a semi-structured interview that captured their experience of the H-H and H-R painting activities.
%~\cite{geiger2024elaborating}.

\subsection{Experimental Setup}
A classroom in the University of Michigan was equipped with curtains to recreate an environment as close as possible to an artist’s studio, as shown in Figure~\ref{fig:conditions}. As artists entered the room, they found a vertical easel, a selection of paints, canvases, brushes, and cleaning supplies. In the H-R experimental sessions, the robot was already present in the room, with the color settings described in Section~\ref{artfram}. Artists were given full creative freedom and were therefore allowed to modify the colors initially proposed by the robot. However, they were not allowed to alter the robot's behavior. The only actions available to them were to stop and restart the robot painting actions. This enabled them to work on the canvas independently (if they wanted), change brushes, or take breaks as needed.

% \begin{figure}[htbp] 
%     \centering  % Center the image
%     \includegraphics[width=0.4
%     \textwidth, scale=0.10]{Figures/H-H.jpg}  
%     \caption{Experimental Set-Up for human collaboration sessions}  
%     \label{setupH}  
% \end{figure}

% \begin{figure}[htbp] 
%     \centering  % Center the image
%     \includegraphics[width=0.4
%     \textwidth, scale=0.10]{Figures/H-R.jpg}  
%     \caption{Experimental Set-Up for robot collaboration sessions}  
%     \label{setupR}  
% \end{figure}

%\\Following each one-hour interaction, participants are asked to complete a personal Artistic Diary. These diaries are designed to capture the participant’s emotional experience during the creative process. Participants are encouraged to reflect freely on their thoughts and feelings, with no predefined questions or prompts. 
%Participants are also encouraged to include visual elements, such as sketches or notes, that may help convey their experience. Diaries are photocopied for research purposes, and the originals are returned to the participants. 

%After the last interaction, we conducted a semi-structured interview lasting 30 minutes with artists, an approach also used by \cite{qin2025encountering} with artists working in artistic collaboration with robots.

\begin{table}[t]
    \scriptsize
    \centering
    \caption{Demographics of Participants.}
    \label{tab:participants}
    \begin{tabular}{cccccc}
        \toprule
        \textbf{ID} & \textbf{Age} & \textbf{Gen.} & \textbf{Style} & \textbf{Exp. (yrs)} & \textbf{Paired} \\
        \midrule
        A1 & 61 & F & Abstracted Realism & 30 & A3 \\
        A2 & 70 & F & Abstract & 25+ & A4 \\
        A3 & 27 & NB & Fantasy mixed media illust. & 12 & A1 \\
        A4 & 63 & M & Interdisciplinary & 40+ & A2 \\
        A5 & 63 & M & Abstract and Figurative & 20+ & A6 \\
        A6 & 61 & M & Abstract and Figurative & 30 & A5 \\
        A7 & 34 & M & Illustrative and Abstract & 20 & A8 \\
        A8 & 65 & F & Figurative and Impressionistic & 50+ & A7 \\
        \bottomrule
    \end{tabular}
\end{table}

\subsection{Measures: Semi-Structured Interview}
After the last session of the experiment, participants took part in a semi-structured interview aimed at exploring their subjective experience, reflecting on the collaborative process, and their perceptions of the robot's role in artistic creation.
%The interview included both demographic/background questions and open-ended items related to the co-creative experience. 
Interviews were conducted in person and audio-recorded for transcription and later qualitative analysis. All transcriptions and interview questions (Appendix\_b) are available on OSF \footnotemark[\value{footnote}].

The semi-structured interview included two primary sections: 1) Demographic and experiential background, and 2) Open-ended reflections on the co-creative painting sessions.
In the first section, participants provided information on their age, artistic background (e.g., type of art, years of experience, primary location), and prior experiences with robots and artistic collaboration. 
The second section focused on participants' subjective experiences during the co-creative process with the robot. Rather than strictly adhering to predefined theoretical constructs, the questions were designed to elicit rich, open-ended responses related to collaboration, agency, creativity, communication, and technological expectations~\cite{adams2015conducting}. 

%The study employs several instruments to capture different aspects of the participants’ experience and the interaction with the robot:
%\subsubsection{Video Recordings} All collaborative sessions are video recorded for subsequent behavioral coding. These recordings allow analysis of interaction dynamics such as turn-taking, imitation, nonverbal communication, and changes in engagement or familiarity over time.
%\subsubsection{Artistic Diaries} After each session, participants complete an open-ended diary to describe their emotional experience during the interaction. Participants may also include sketches or notes. Diaries provide qualitative, introspective data on feelings, reflections, and evolving relationships with the robot or human partner. The following instruction is included in each diary:

%\textit{Please take your time to describe how you felt during the drawing interaction. You can use stream-of-thought, you can draw, paint, write a poem, or you can use any other way to express yourself. There are no right or wrong answers, and you are not required to share anything you do not wish to. Your honest reflections, feelings, and impressions are entirely welcome. Feel free to express yourself in whatever way feels most comfortable to you.}

\section{Qualitative Analysis}
We conducted a reflexive thematic analysis ~\cite{braun2019reflecting} of the eight interviews. The thematic structure informed by the four main research questions shaped the analytical focus. While the analysis remained open to unexpected insights, the coding process was deductive. The analytic process followed the key phases: we began with familiarization through repeated reading of transcripts, followed by initial coding aligned with each research question. Codes were reviewed and grouped into broader thematic categories that reflected shared meanings across participants. Final themes were refined to ensure internal coherence and clear distinctions between categories.
All transcripts (N = 8) were analyzed using MAXQDA software.
%~\cite{kuckartz2019analyzing} %\footnote{\url{https://www.maxqda.com/}} software. %A single researcher conducted the coding process, in line with reflexive thematic analysis practices, which emphasize depth of engagement and analytic reflexivity over inter-coder reliability. 
This approach enabled us to identify five main themes: (1) Comparing Human-Human and Human-Robot Collaboration in Painting Production, (2) Perceived Agency of the Robot, (3) Creative Flow, (4) Robot's Capabilities: Social Interaction, and (5) Robot's Capabilities: Painting Techniques.

% \begin{figure}[t]
%     \centering
%     \includegraphics[width=\linewidth]{Figures/exp setup.jpg}
%     \caption{\small{Study set-up: (a) human-human artist collaboration; (b) human-robot collaboration. The two configurations of the studio were created in the same room.}}
%     \label{fig:set-up}
% \end{figure}

\section{Results}

\subsection{\textbf{Theme 1. Comparing Human-Human and Human-Robot Collaboration in Painting Productions}}
The comparison between H-H and H-R collaboration revealed distinct experiential differences. H-H interactions were often described as intuitive and socially engaging, while H-R collaborations felt more reflective. Notably, some participants appreciated the robot's lack of emotional demands, noting it allowed them greater creative freedom.

\textbf{Co-presence and human exchange} emerged as a key aspect of H-H collaboration. Participants emphasized the value of communication with their human partner, which facilitated mutual understanding and fostered a sense of intimacy. As A8 explained, \textit{``With A7, it was more personal because as we painted, we talked and we got to know each other. So it was a more friendly environment, whereas with the robot, it's kind of sterile.''} Similarly, A2 recalled, \textit{``What I most remember and enjoy is getting to know him [referring to A4] through the painting process.''} Even when stylistic differences emerged, they were negotiated through human sensitivity and mutual respect, A5: \textit{``It was an adjustment. It was fun. I felt like I had to... I wanted to... respect the other artist. [...] We were both careful about that, but we had very different styles.''}.

In contrast, \textbf{anticipation and strategy with the robot} were central to the H-R interaction. Participants described the need to observe the robot and anticipate its behavior. As A8 noted, \textit{``I observed it first and tried to pick up some rhythms from it. [...] I was trying to anticipate what it was gonna do.''} 

Some participants articulated more \textbf{reflective and anticipatory modes} {of engagement during the human–robot collaboration. Rather than reacting spontaneously to another person’s gestures, artists described pausing to observe the robot’s behavior, thinking ahead about how its next action would be, and adjusting their own interventions accordingly. This form of reflection was less socially oriented and more self-directed, involving prediction about when and how to intervene on the canvas. As A2 noted, \textit{“With the robot, I was thinking a lot before interacting with it,”} highlighting a shift from interpersonal coordination toward a more introspective and deliberative creative process.
Finally, participants frequently referred to \textbf{creative freedom and negotiation} when comparing the two experiences. Some participants felt freer working with the robot, as they did not have to manage interpersonal dynamics or consider the potentially conflicting perspectives of another human artist. This made the interaction feel less intimidating and more personally expressive. The robot was sometimes described as a \textbf{neutral tool that encouraged creativity}, as A1 mentioned: \textit{``With the robot, it was more like you having a robotic assistant doing things that also pushed your creativity a bit.''} For others, this lack of negotiation opened space for full control over the creative process. A5, for example: \textit{``Maybe I felt freer to be completely in control, because the robot was making random marks, so it was going to be up to me to make the work interesting. Whereas with the other artist, we were both trying to show mutual respect and not dominate one another, so maybe that inhibited the creativity a little bit.''}. Although the robot’s strokes were generated according to a predefined artistic framework, this structure was not directly legible to A5 during the interaction. From the artist’s perspective, the robot’s gestures often appeared unscripted. As a result, these marks were interpreted as “random,” not because they were unstructured, but because the rationale guiding the robot’s actions remained inaccessible within the interaction.

\subsection{\textbf{Theme 2. Perceived Agency of the Robot}}
Participants reflected on the role of the robot in the artistic process, oscillating between a passive tool or an autonomous collaborator. 
\textbf{Balancing control and autonomy} was a recurring topic in the interviews. Several participants expressed a desire for a mixed-mode collaboration in which both the human and the robot could initiate actions. As A8 remarked, \textit{``I would like to give the robot the opportunity to choose [what to do] and then I would choose as well.''} Similarly, A2 noted, \textit{``If I could negotiate with the robot to start and stop, or do this or do that, I could imagine what I might do.''} 
%Others leaned more strongly toward maintaining control, preferring a robot that would follow explicit commands: \textit{``I think I would prefer a robot to do what I wanted it to do,''} stated A$3$.

Interest in \textbf{reciprocal interaction and initiative} was common. Some participants wanted the robot to not only respond to input but also to initiate actions or offer suggestions. A3 imagined a scenario where the robot could act like a peer: \textit{``Sometimes it's nice to have somebody pick one [i.e. artistic piece]. I could just hold them up to the robot [...] it could point to one.''}. A1 expressed curiosity about a robot that could interrupt routines to challenge artistic habits: \textit{``If I was getting complacent, maybe have it decide to switch things up.''}

\textbf{Metaphorical framings} were often used to describe the robot's actions and role. A3 stated: \textit{``It was like working with a cart on a track.''}. A$7$ instead expressed: \textit{''I would compare it to drawing outside and the wind blowing, like drawing a tree and the leaves moving.''}
Notably, some artists paralleled the experience as painting with children, like A$4$ \textit{''it reminded me when I was painting with my daughters when they were very, very young age. Just observing." } and A$8$ \textit{``It's making brushstrokes and choosing top or bottom, like when a child paints.''}. The connection with child-like experiences denotes the non-judgmental space held in H-R interactions compared to H-H interactions. Additionally, it also denotes a robot that is technically more limited than a human in its painting technique.

\begin{figure*}[!t]
    \centering
    \includegraphics[width=\textwidth]{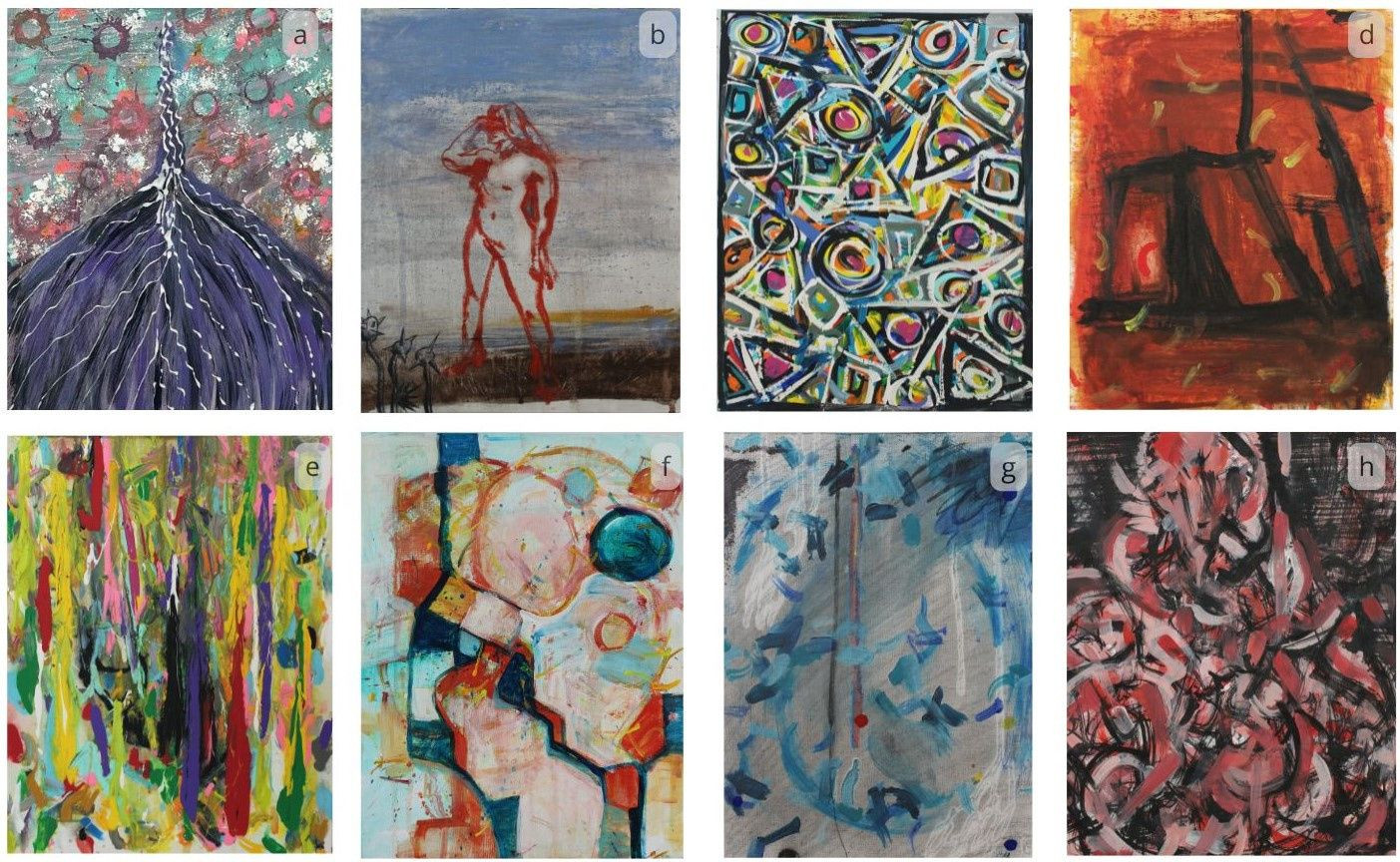}
    \caption{\small{Artworks produced by artist collaboration (a and b) and by collaboration with the robot (c-h) during this study.}}
    \label{fig:paintings}
\end{figure*}

\subsection{\textbf{Theme 3. Creative Flow}}

Some participants described experiences that we interpret as aspects of a flow-like creative state during the sessions with the robot, including moments of \textbf{inspiration and generative engagement, as well as surprise and frustration.
}. For some, the interaction with the robot disrupted their usual creative rhythm, requiring adaptation and patience. As A7 noted, \textit{``I found myself a bit frustrated by its decisions [...] maybe it's just because I didn't understand what it was doing''}. A4 pointed out: \textit{{``It was unhelpful, mostly unhelpful [...] trying to adapt to something you have basically no communication with.''}}. 
However, a subset of participants described moments of sustained creative engagement that contrasted with earlier feelings of disruption or frustration.
A2 reflected, \textit{``There was just enough structure but not too much structure in the mark-making so that I could imagine covering it, extending the lines, playing with the color, all sorts of things''}.
This description suggests a balance between constraint and openness that supported ongoing exploration and uninterrupted creative action.

Several participants emphasized that the robot, while not creative in itself, acted as a catalyst for their own creativity. As A1 put it, \textit{``It really inspired me to work differently than I normally do, to respond to things I wouldn't usually respond to''}. A1 further described the robot as \textit{``a complementary person that was saying, `why don't you try this?' instead of me just doing what I normally do.''} A6 described the process as both \textbf{disruptive and generative}: \textit{[the robot] ``Integrated and disrupted at the same time, because that's what creative process is''}. Others found the robot's limitations to be a source of creative tension. A5 explained, \textit{``It was mostly for me just trying to make an interesting work of art while combating sort of randomness''}. It is interesting to note that while many participants did not ascribe creative skills to the robot, they recognized a sense of its creative style. 

\subsection{\textbf{Theme 4. Robot capabilities: Social Interaction}}
Participants expressed a range of expectations regarding how the robot should \textbf{learn and adapt within an interaction} during artistic collaboration. 
%\textbf{Learning and adapting within an interaction} appeared in subtle forms. 
They frequently mentioned adjusting their speech or expectations to better collaborate with the robot. A1 acknowledged, \textit{``I figured out how it would react better if I spoke more clearly.''} 

The \textbf{desire for expressive and multimodal communication} with the robot was consistent. Several participants envisioned the robot using not just speech, but also non-verbal cues to enhance collaboration. As A8 noted, \textit{``I guess [...] music. Because music enhances the mood.''} Others proposed technical augmentations, such as pointing mechanisms or gesture-like behaviors. A7 suggested, \textit{``It would be nice to have it indicate what it's seeing [...] like pointing a laser at an area.''}

Participants reported \textbf{mixed comfort with verbal interaction}. While some welcomed the idea of the robot talking, finding it potentially enriching: \textit{``That would be great. I would love it if it talked to me,''} stated A1. Others, however, found the idea less appealing as A7 explained, \textit{``That would, for me personally, be unsettling.''}
%\Textbf{Expectations for integrating user-provided materials} also surfaced. Several artists were open to the robot referencing sketches, photos, or prior artworks. A2 stated, \textit{``Yeah, you can show it pictures [...] and you can start going from that.''} Others, like A5, were more hesitant: \textit{``I don't know if I would like that or not. [...] I wouldn't want it to necessarily imitate what I do.''} 

\subsection{\textbf{Theme 5. Robot's Capabilities: Painting Technique}}
Participants engaged both critically and imaginatively with the robotic painting system, acknowledging its \textbf{mechanical precision alongside artistic limitations}—often using this contrast as a starting point for envisioning more co-creative possibilities.
The most frequently mentioned limitation was the robot's \textbf{restricted and repetitive movement}. Participants were generally expecting more from the robot's capacity and noted its gestures were often predictable and lacked organic variation. As A8 stated, \textit{``It was limited because it's just brush strokes. [...] It didn't do a continuous line or [...] organic shapes.''} Similarly, A3 remarked, \textit{``It moves along this certain track [...] the marks themselves are fairly predictable: curve, curve, curve, curve.''} Others, like A4, emphasized how little motion was involved: \textit{``The robot has this motion, very little motion.''} or also, \textit{''Then I realized, okay, the robot has a very limited motion. It goes six times into the canvas. It puts six brush strokes. They could be different brush strokes. And then the robot goes back, cleans up to use another color and does the same process again and again, but in a different part of the canvas.''} It is interesting to note how the artists carefully studied the robot's actions and dynamics of painting to be able to predict and collaborate with it.

The robot's \textbf{limited expressive range} was a recurring critique. As A5 commented, it was: \textit{``very sophisticated in its mechanical movements, but really primitive in its mark-making,''}.
Despite these constraints, participants took this opportunity to imagine what a more advanced robot could do. A recurring idea was to \textbf{expand the robot's technical and material capabilities}, envisioning a system that could handle not only brushes but also pencils, markers, palette knives, or airbrushes. As A8 put it, \textit{``I would like for it to be able to not only just handle a brush, but to use other instruments as well.''} A3 expressed interest in modular customization: \textit{``I would absolutely love for it to have 20 more attachments.''} 

Control over \textbf{mark-making parameters} such as pressure, direction, and speed was also seen as critical for enhancing creative expression. A1 emphasized, \textit{``Pressure is really important in a lot of ways. You can tell it, I want you to really be light with this area or I want you to really press into this.''} Others, like A2, desired a broader range of strokes, including circular, sweeping, or calligraphic gestures. Another frequent aspiration was for more \textbf{adaptive and responsive behavior}. A5 imagined the robot equipped with computer vision that could interpret the canvas and adjust accordingly, envisioning a system that could \textit{``help me achieve the optimum amount of contrast.''} Some imagined a more advanced robot as a smart studio assistant, able to work on backgrounds or repetitive tasks overnight (A4, A6), while others hoped it could emulate specific artistic styles or respond to reference materials, as A2 mentioned, perhaps the robot \textit{``Could you do the background like the artist Monet?''}

While it's interesting to consider the robot as a historically and contextually aware agent, this also raises the question of copyrights. Would artists, and the society at large, really want a robot that can copy another artist's style?  

\begin{figure}[t] 
    \centering  % Center the image
    \includegraphics[width=\linewidth]{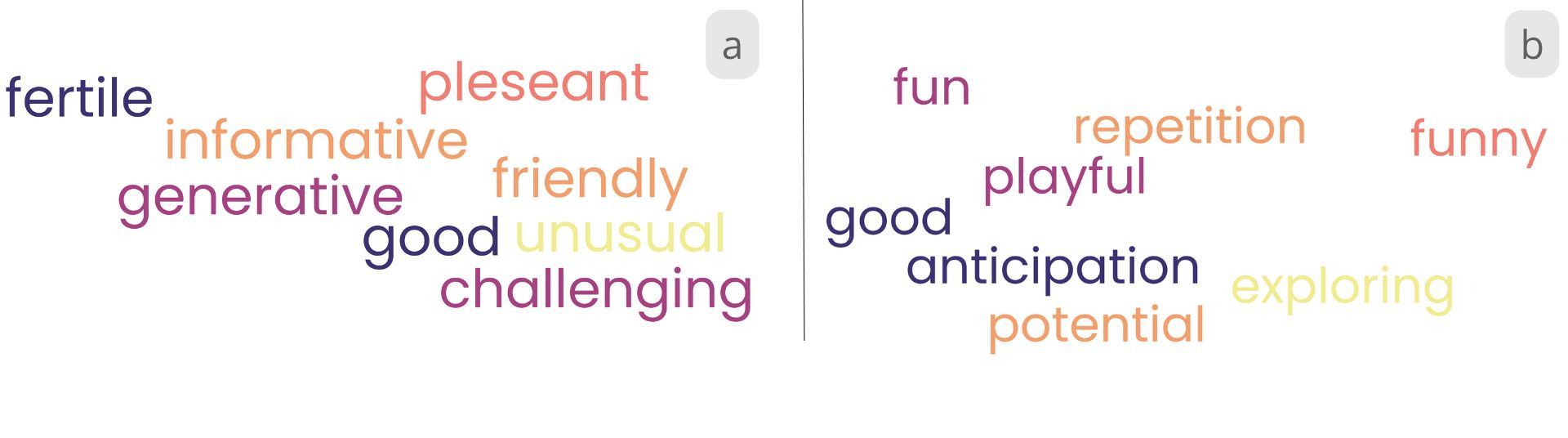}  
    \caption{\small{Outcomes of the single-word to describe the interaction sessions: (a) words for human-human collaboration; (b) words for human-robot collaboration.} } 
    \label{Fig:wordcloud}  
\end{figure}

\subsection{\textbf{Word Cloud}}
During the interviews, participants were asked to describe the interaction with the other artist using a single word, and to do the same for the interaction with the robot. The results are reported in the Figure \ref{Fig:wordcloud}. Mirroring the results that emerged from the thematic analysis, the concepts in the wordcloud reflect how H-H collaboration appears more \textbf{cognitively and socially engaging}, while H-R collaboration is framed as \textbf{playful and explorative}.
This positivity does not negate the frustration and limitations expressed towards the robot, but rather coexists with them, reflecting different layers of the experience captured by single-word and narrative responses.

\section{Discussion}
This exploratory study offers a rich, situated exploration of how professional artists perceive and experience collaboration with an autonomous painting robot, compared to working with another human artist. Through thematic analysis of semi-structured interviews, we identified challenges and opportunities of human-robot artistic co-creation. These findings contribute to an emergent understanding of creative \ac{HRI} that moves beyond output quality to consider relational, affective, and cognitive dynamics.

One of the most striking findings is the different nature of \textbf{collaboration} experienced in the two conditions. H-H sessions were often described as socially rich, intimate, and dialogic, facilitated by mutual adaptation. This aligns with prior research showing how synchrony and interpersonal interaction enhance rapport in joint artistic activities~\cite{abraham2023painting, rabinowitch2017synchronized}. In contrast, H-R collaborations were experienced as more reflective, self-driven, and structured, but also playful. While the robot lacked emotional responsiveness, this absence also allowed some participants to feel greater freedom and control over the creative process, and eventually even feel less intimidated and experience more fun. The robot was often seen as a playful, neutral partner, removing the social pressures that could sometimes emerge in H-H collaboration. 

A key theme is the ambivalent perception of the \textbf{robot’s agency}. Some artists viewed it as a mere tool, others as a more autonomous collaborator. Several participants expressed a desire for more fluid agency-sharing, envisioning systems where both human and robot could take initiative or alternate control. Others preferred retaining clear control and saw autonomous behaviors as intrusive. This diversity of preferences suggests the value of flexible, mixed-initiative systems where levels of autonomy can be adapted to the user.

Participants also employed metaphors to make sense of the robot's role. These framings illustrate how H-R artistic interaction is not only functional but deeply interpretive: artists are continuously trying to ``read'' the robot's behavior and situate it within their own meaning-making processes. Notably, a similar metaphor emerged in the collaborative painting experience between the Alter-Ego robot and artist Melkio, who described the interaction \textit{``as if I were painting through a 3-year-old child.''} This parallel highlights how the child-like framing may capture both the perceived playfulness of the robot's contributions and its limitations. The dimension of the ``game'' or `fun'' in the H-R collaboration is also evident from the wordcloud results.

The impact of the robot on artists' \textbf{creative flow} was multifaceted. %\hl{Even if participants did not explicitly label these experiences as “flow”, the term is used here to denote partial and situated qualities of engagement observed in the narratives, which coexisted with moments of difficulty and perturbation.}
For some, the robot disrupted their usual rhythm, requiring patience, adaptation, and new strategies. For others, it became a source of surprise and inspiration. The structured behavior of the robot challenged conventional patterns and sometimes led to unexpected aesthetic decisions. These accounts support recent perspectives that emphasize disruption, uncertainty, and constraint as generative forces in creativity, similar to what was found previously ~\cite{qin2025encountering, he2023exploring}. Even when the robot was perceived as limited or repetitive, several artists described moments of improvisation and transformation that arose precisely because of these constraints. 

Lastly, artists expressed a broad range of expectations for how the robot could support \textbf{more interactive and expressive collaboration}. These included desires for multimodal communication, contextual responsiveness, and adaptability to the evolving artwork. 
%Some participants imagined robots that could understand sketches or reference images, suggest next steps, or provide feedback through visual or verbal cues. 
%This suggests that future systems should go beyond command-and-control architectures and instead support rich, continuous co-creative dialogue between human and machine.

\section{Conclusions}
This exploratory study explored how professional abstract artists perceive and experience collaboration with an autonomous painting robot, in direct comparison with human artistic collaboration. We identified nuanced perspectives on creative agency, flow, social presence, and the perceived limitations and potentials of robotic co-creation.
Our findings suggest that while H-R collaboration lacks the emotional reciprocity and richness of H-H interaction, it can nevertheless serve as a productive space for reflection, inspiration, experimentation, and creativity.
These qualities coexisted with clear technical limitations and often reflected artists’ adaptive reinterpretations rather than intrinsic features of the robot.
\\Artists were not merely passive users of the robotic system: they actively constructed meaning around the robot's actions, incorporated its gestures into their practice, and imagined future enhancements that would allow for more expressive, dialogic interactions.
The insights from our research can meaningfully inform future implementations of robotics in art. By relying on the desires and needs reported by these artistic stakeholders, we can support more attuned technological developments in the artistic domain.

This work represents one of the first attempts to foreground the perspectives of domain experts as key stakeholders in the development of autonomous robotic systems for the arts. It also stands out by adopting a long-term engagement approach, allowing participants to interact repeatedly with the system over time, rather than relying on one-off encounters, as is common in \ac{HRI} studies. Finally, this exploratory work embraces a multidisciplinary framework that integrates artistic practice, robotics, and qualitative research methods, contributing to an increasingly diverse and ecologically valid vision of \ac{HRI}.
%\\By centering the voices of experts and bridging disciplinary boundaries, we move closer to realizing robots not merely as tools or novelties, but as meaningful partners in the evolving landscape of creative collaboration.

\section*{Acknowledgments}
This project is funded by DARPA Young Faculty Award (YFA) \#D24AP00323-00. The content of this research is solely the responsibility of the authors and does not necessarily represent the official views of DARPA. We are grateful to Angshu Adya, Cindy Yang, David Ho, Emily Wu, Hana Ichikawa, Longzhen Yuan, Rishad Hasam, and Tianxin Li for their contributions to the robot system. We thank the artists who participated in this study for their valuable time and insights. 



\bibliographystyle{IEEEtran}
\bibliography{biblio}

\newpage

\vfill

\end{document}